\documentclass[conference,compsoc]{IEEEtran}

\usepackage[breaklinks,colorlinks, citecolor=black, urlcolor=black, linkcolor=black, citecolor=black]{hyperref}
\usepackage{color}
\usepackage{amsmath,amsopn}
\usepackage{amssymb}
\usepackage[ruled,boxed,commentsnumbered, linesnumbered]{algorithm2e}
\usepackage[labelformat=simple]{subcaption}

\usepackage{endnotes,microtype,xspace,graphicx,fancyvrb,multirow}
\usepackage{booktabs}
\usepackage{array,underscore,relsize}
\usepackage[T1]{fontenc}
\usepackage{fancyhdr,totpages}
\usepackage{enumitem}
\usepackage[font=small, labelfont={bf}]{caption}
\usepackage{tabularx}
\usepackage{colortbl}

\usepackage{tcolorbox}
\usepackage[sort]{cite}

\usepackage{fp}
\usepackage{siunitx}
\usepackage{diagbox}

\usepackage[frozencache,cachedir=.]{minted}
\AtBeginEnvironment{listing}{\setcounter{listing}{\value{lstlisting}}} 
\AtEndEnvironment{listing}{\stepcounter{lstlisting}}

\usepackage{balance}

\definecolor{lightgray}{rgb}{0.91, 0.91, 0.91}
\definecolor{lightblue}{rgb}{0.8627, 0.9176, 0.9686} %

\usepackage{multicol}

\usepackage{verbatim}
\usepackage{lipsum}

\usepackage[normalem]{ulem}

\newcommand{\sys}{\mbox{\textsc{Neo}}\xspace}

\newcommand{\enigma}{\mbox{EnIGMA}\xspace}

\newcommand{\numapp}{25\xspace}
\newcommand{\numtestapp}{21\xspace}
\newcommand{\numgroundapp}{4\xspace}

\newcommand{\numgroundappvul}{18\xspace}

\newcommand{\numlang}{7\xspace}

\newcommand{\numnewvulcorpus}{22\xspace}
\newcommand{\numnewvulbench}{2\xspace}
\newcommand{\numnewvul}{24\xspace}
\newcommand{\numnewvulmore}{18\xspace}
\newcommand{\numnewvultotal}{42\xspace}

\usepackage[dvipsnames,table]{xcolor}

\usepackage{url}
\newcommand{\cc}[1]{\mbox{\smaller[0.5]\texttt{#1}}}

\fvset{fontsize=\scriptsize,xleftmargin=8pt,numbers=left,numbersep=5pt}

\def\Snospace~{\S{}}

\newif\ifdraft\drafttrue
\newif\ifnotes\notestrue
\ifdraft\else\notesfalse\fi

\newcommand{\eg}{{\em e.g.}}

\newcommand{\etc}{{\em etc.}\xspace}
\newcommand{\ie}{{\em i.e.}}

\input{glyphtounicode}
\newcommand{\squishlist}{
\begin{itemize}[noitemsep,nolistsep,leftmargin=10pt]
  \setlength{\itemsep}{-0pt}
}
\newcommand{\squishend}{
  \end{itemize}
}

\usepackage{tikz}

\usepackage{xstring}
\newcommand{\PP}[1]{
\noindent{\bf #1.}
}

\newtcolorbox{mybox}[3][float=ht]
{
  colback=#2!5!white,
  colbacktitle=#2!15!white,
  boxrule=0.25mm, 
  top=0pt, bottom=0pt, left=0pt, right=0pt,
  coltitle=black,
  title    = {#3},
  #1,
}

\newcounter{prompt}

\newcommand{\summarybox}[1]{%
    \begin{tcolorbox}[colback=blue!3!white, boxrule=0.25mm, sharp corners, colframe=black, top=0pt, bottom=0pt, left=0pt, right=0pt]
    \textbf{Summary:}#1 %
    \end{tcolorbox}
}

\newcommand{\appautoref}[1]{\hyperref[#1]{Appendix~\ref{#1}}}
\newcommand{\algcomment}[1]{\tcp*[h]{\textcolor{blue}{\scriptsize{#1}}}}

\begin{document}

\title{Detecting Privilege Escalation in Polyglot Microservices via Agentic Program Analysis}

\author{
\IEEEauthorblockN{Penghui Li}
\IEEEauthorblockA{Columbia University\\
pl2689@columbia.edu}
\and

\IEEEauthorblockN{Hong Yau Chong}
\IEEEauthorblockA{Columbia University\\
hc3661@columbia.edu}
\and

\IEEEauthorblockN{Yinzhi Cao}
\IEEEauthorblockA{Johns Hopkins University\\
yinzhi.cao@jhu.edu}
\and

\IEEEauthorblockN{Junfeng Yang}
\IEEEauthorblockA{Columbia University\\
junfeng@cs.columbia.edu}

}

\date{}
\maketitle

\begin{abstract}

Microservices are widely adopted in modern cloud systems due to their scalability and fault tolerance.
However, microservice architectures introduce significant complexity in privilege and permission control, creating risks of privilege escalation where attackers can gain unauthorized access to resources or operations.
Detecting such vulnerabilities is challenging due to complex cross-service interactions, polyglot codebases, and diverse privileged operations and permission checks. We present \sys, an agentic program analysis framework that combines large language models (LLMs) with classic program analysis to address these challenges.
\sys leverages an LLM-based agent that dynamically generates analysis plans, adapts code search strategies, and validates semantics.
We develop code search primitives that enable \sys to perform scalable and flexible code exploration across services and languages.
We evaluated \sys on \numapp open-source microservice applications spanning 7 programming languages and 6.2 million lines of code. \sys uncovered \numnewvul zero-day privilege escalation vulnerabilities and achieved 81.0\% precision and 85.0\% recall on a ground-truth dataset. Compared to existing program analysis and agentic solutions, \sys demonstrated significant improvements in both detection accuracy and scalability. We further showcased \sys's extensibility by applying it to other application domains and vulnerability types, uncovering \numnewvulmore additional zero-day vulnerabilities.

\end{abstract}

\section{Introduction}
\label{s:intro}

Microservice architectures have become the dominant paradigm for building large-scale distributed systems in the cloud~\cite{microservices}.
By decomposing monolithic applications into loosely coupled, independently deployable services, microservices enable high scalability, fault tolerance, and rapid development~\cite{newman2019monolith, abgaz2023decomposition}.
Major companies like Amazon~\cite{amazon-microservices-evolution} and Google~\cite{google-ebay-microservices} have evolved their internal systems to use microservices, often deployed in polyglot architectures.
They also provide cloud platforms (\eg, AWS, Google Cloud) that enable other organizations to adopt microservice architectures~\cite{aws-microservices, google-microservices}.

However, the microservice architecture makes permission control inherently complex.
On the one hand, modern microservice systems involve numerous principals, such as users, roles, groups, service accounts, \etc
On the other hand, each service manages a wide range of service-specific resources and implements diverse privileged operations using different programming languages and frameworks.
Authentication (authN) and authorization (authZ) checks for permission control are distributed across services.
All these make it extremely challenging to ensure privileged operations are properly protected.

\begin{figure}[h]
\center{\includegraphics[width=0.95\columnwidth, trim=10pt 0 0 0, clip]{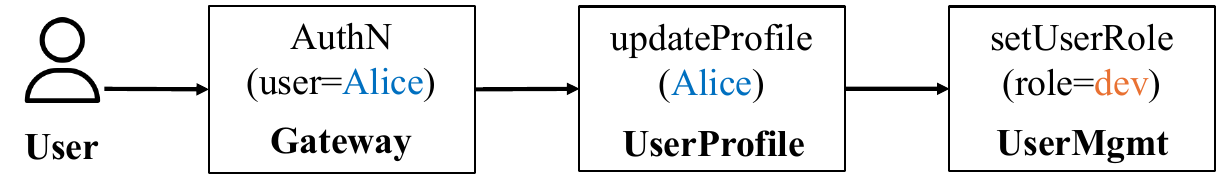}}
\caption{A user role update flows through three services.}
\label{fig:privesc-example}
\end{figure}

\PP{Motivating Example}
We illustrate this complexity using a real-world scenario in \autoref{fig:privesc-example}.

When the user Alice initiates a profile update to switch to the \cc{developer} role, the request flows through three services.
The Gateway service acts as the system’s entry point and performs authentication by verifying Alice’s credentials and routing her request to the appropriate backend.
The UserProfile service handles user-specific updates and prepares the role-switching request.
Finally, the UserMgmt service manages user accounts and executes the \cc{setUserRole(role=dev)} operation to finalize the change.
Beyond authentication, this operation actually requires authorization to verify whether Alice is eligible for this role switch.
Such a check could be implemented in multiple places: at the Gateway before forwarding the request, at the UserProfile before invoking UserMgmt, or at the UserMgmt before executing the role change.
In implementation, each service must decide which checks it should perform and which it can assume other services have already validated.

When services fail to coordinate on authentication or authorization responsibilities, privilege escalation vulnerabilities arise.
For instance, even if an authorization check validates that Alice can perform role changes, without verifying which specific roles she can assume, an attacker can manipulate the request to specify \cc{role=admin} instead of \cc{role=dev}.
In fact, this is a realistic vulnerability we found in Mall4cloud~\cite{mall4cloud}, a popular cloud-based application with around 6K stars on GitHub.
Such privilege escalations have been known with severe consequences, such as remote code execution~\cite{crayfish-rce-cve}, gaining root privileges~\cite{cisco-microservices-cve}, and compromising entire infrastructure~\cite{mspolicy}.

Systematically detecting such privilege escalation vulnerabilities in polyglot microservice systems presents multiple challenges.
The first challenge lies in the inherent complexity of distributed, heterogeneous microservice systems (\textbf{C1}).
As illustrated earlier, a single request often spans multiple services that are implemented using diverse programming languages and frameworks, and these services communicate through different protocols (\eg, gRPC~\cite{grpc}, REST~\cite{rest-api}, Kafka~\cite{kafka}).
No individual service has a complete view of the end-to-end privilege enforcement logic, and thus requires analysis to holistically reason across service boundaries.

Second, it requires deep semantic understanding beyond structural code patterns (\textbf{C2}).
Unlike memory corruptions (\eg, buffer overflows) that can be characterized by code structure, finding privilege escalation requires recognizing which operations are privileged and which code implements authN/authZ checks.
For example, determining that \cc{setUserRole()} is a privileged operation demands semantic understanding of the function's purpose and security intent.
Traditional program analysis approaches~\cite{mace, son2011rolecast, pcfinder, mocguard, mpchecker} rely on security experts to manually summarize such patterns and do not scale to diverse microservice implementations.

Third, even after identifying privileged operations and authN/authZ checks, it remains challenging to correlate which checks actually protect a given operation (\textbf{C3}).
This requires analyzing the control and data dependencies between the operation and potential checks.
These dependencies can traverse diverse structures ranging from simple conditional statements and function call sites to framework-specific mechanisms like decorators, middleware chains, or policy modules.
For instance, determining what protects an endpoint may require tracing from the endpoint to a decorator, then to a router definition, and finally to the check implementation.
Each application structures these relationships differently, where a fixed analysis pipeline does not suffice.

To the best of our knowledge, no prior work can solve these challenges.
Existing microservice analysis tools~\cite{mscan, microscope, mspolicy} focus on other concerns and do not detect privilege escalation vulnerabilities.
For example, MScan~\cite{mscan} targets taint-style vulnerabilities such as command injection in Java-based microservices but does not reason about privileges.
Microscope~\cite{microscope} analyzes how code changes in one service impact others during development, but does not perform any security analysis.
On the other hand, privilege escalation detection approaches developed for monolithic web~\cite{mace, son2011rolecast} and mobile~\cite{davi2010privilege} applications cannot handle polyglot microservices because they assume a single-language codebase with no cross-service interaction.

Our key insight is that large language models (LLMs) and classic program analysis have complementary strengths. LLMs excel at semantic reasoning about code intent, while classic program analysis provides scalable cross-service tracking. Combined through an agentic design, they enable dynamic correlation across complex code structures.
For \textbf{C1}, we design code search primitives that abstract classic static analysis operations (\eg, flow tracking, call-graph traversal) with language-agnostic interfaces using CodeQL~\cite{codeql}, enabling scalable cross-service analysis across polyglot codebases.
For \textbf{C2}, we leverage LLMs to interpret semantic cues from program identifiers and natural language artifacts (\eg, function names, comments) to identify privileged operations and reason about security check adequacy.
For \textbf{C3}, an LLM agent uses these primitives to trace control and data dependencies, and correlate operations and checks across different structural patterns.

We realize these ideas in a novel agentic program analysis framework, called \sys.
\sys iteratively orchestrates code search primitives and LLM reasoning to identify privileged operations, trace cross-service flows, and locate and validate authN/authZ checks.
A key novelty of \sys is that it treats code as structured program representations and leverages code search primitives to enable efficient, on-demand context retrieval from large polyglot codebases.
Prior code agents~\cite{sweagent, openhands, enigma} rely on LLMs to directly read and reason about entire files as unstructured text, which limits scalability and easily exhausts the context.
While implementing our primitives atop CodeQL~\cite{codeql} inherits its inherent language-modeling limitations, this foundation ensures structural precision and cross-service scalability that are currently infeasible through purely text-based agent reasoning.

We extensively evaluated \sys on \numapp open-source microservice applications, including an evaluation corpus of \numtestapp applications and a ground-truth dataset of \numgroundapp applications.
\sys successfully identified \numnewvul zero-day privilege escalation vulnerabilities.
On the ground-truth dataset, \sys achieves 81.0\% precision and 85.0\% recall.
Compared to prior classic and agentic program analysis solutions, \sys demonstrates significant improvements in detection accuracy, scalability, and vulnerability coverage.
For instance, compared to the \enigma agent that directly reads and analyzes entire source files, \sys found 24 more vulnerabilities.
Our ablation study further shows that the code search primitives are essential: without them (\eg, directly invoking CodeQL for context retrieval), the system missed most of the vulnerabilities.
Beyond privilege escalation, we demonstrated \sys's extensibility by applying it to detect other vulnerability types across different application domains, uncovering \numnewvulmore additional zero-day vulnerabilities.
All vulnerabilities were responsibly disclosed to maintainers, with 8 fixed to date.

\squishlist
\item We designed code search primitives that enable flexible code exploration and context retrieval.
\item We developed \sys, a novel agentic program analysis framework that combines LLM semantic reasoning with scalable program analysis.
\item \sys discovered a total of \numnewvultotal new vulnerabilities and demonstrated superior performance over existing methods.

\squishend

\begin{figure*}[t]
  \begin{minipage}[b]{.53\textwidth}
    \begin{subfigure}{\linewidth}
    \centering
    \inputminted[breaklines=true, frame=single]{Java}{code/example1.java.tex} %
    \caption{UserProfile service written in Java.}
    \label{code:example-java}
    \end{subfigure} %
  \end{minipage}%
  \hspace{10pt}%
  \begin{minipage}[b]{.45\textwidth}
    \begin{subfigure}{\linewidth}
    \centering
    \inputminted[breaklines=true, frame=single, firstnumber=16]{Python}{code/example1.python.tex}
    \caption{UserMgmt service written in Python.}
    \label{code:example-python}
    \end{subfigure}
  \end{minipage}%
\caption{Code implementation of the cross-service privilege escalation vulnerability in \autoref{fig:privesc-example}. Both services perform authentication but fail to validate eligibility for the specific role being requested, allowing authenticated users to escalate to arbitrary roles.}

\label{code:example}
\end{figure*}

\section{Background}
\label{s:background}
\subsection{Microservice Architecture}
\label{s:background-microservice}
Microservice architectures decompose applications into multiple services that communicate over network protocols.
A characteristic of microservices is their \emph{polyglot nature}, where different services can be implemented in different programming languages and frameworks.
For example, a data analytics service might use Python for its rich scientific libraries, while a high-performance API gateway uses Java or Go, and a real-time notification service leverages Node.js for asynchronous I/O.
Services interact with each other through REST APIs~\cite{rest-api}, gRPC~\cite{grpc}, GraphQL~\cite{graphql}, and message brokers like Kafka~\cite{kafka} and RabbitMQ~\cite{rabbitmq}.
While this flexibility enables organizations to choose optimal technologies for each component, it introduces challenges for cross-service security analysis since traditional static analysis tools are typically single-language.

\subsection{Permission Control and Privilege Escalation}
\label{s:background-permission}
Microservice applications often implement two distinct security mechanisms for permission control.
\emph{Authentication (authN)} verifies the identity of the requester by confirming ``who you are'' through credentials like passwords, tokens, or certificates.
\emph{Authorization (authZ)} determines what the authenticated user is permitted to do by verifying ``what you can access'' based on the user's identity and the requested resource.
For example, in the scenario of role-switching, verifying Alice's credentials confirms authentication, but checking whether Alice is permitted to switch to the \cc{developer} role confirms authorization.
\emph{Privilege escalation} occurs when a user gains access to resources or operations beyond what they are authorized to have, such as when security checks are missing, insufficient, or improperly implemented, allowing users to bypass intended access restrictions.

\subsection{LLM-based Code Agents}
\label{s:background-agent}
LLM-based agents are autonomous systems that iteratively perceive their environment, reason about goals, and take actions through tool invocations to achieve objectives.
Recent work has applied this paradigm to code analysis tasks~\cite{openhands, sweagent, enigma, agentless, llift}, where agents interact with codebases through tools like file system operations and command execution.
For example, SWE-agent~\cite{sweagent} provides agents with bash commands to navigate repositories, such as listing directories with \cc{ls}, searching keywords with \cc{grep}, and reading file contents.
Building upon SWE-agent, \enigma~\cite{enigma} extends the agent framework to cybersecurity tasks by introducing interactive tools that enable the LLM to use debuggers and other utilities essential for vulnerability analysis.
These agents operate through a perception-action loop.
The LLM observes the current state (\eg, file contents, execution results), reasons about next steps, invokes tools to gather more information or make changes, and directly processes the returned outputs to guide subsequent decisions.

\begin{figure*}[t]
    \center{\includegraphics[width=0.95\textwidth]{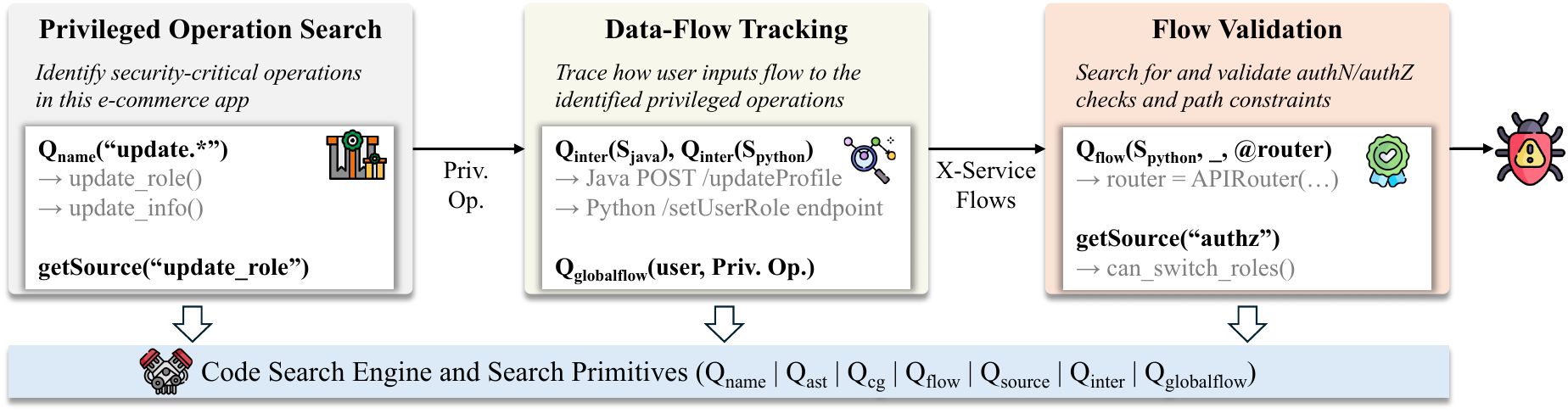}}
\caption{Workflow of \sys. The LLM agent uses code search primitives to iteratively identify privileged operations, trace cross-service flows, and validate security checks.}
    \label{fig:workflow}
\end{figure*}

\section{Motivation}
\subsection{Motivating Example}
\label{s:motivating-example}

We present in \autoref{code:example} simplified code for the vulnerability introduced in \autoref{fig:privesc-example}.
The two services involved in the vulnerability use Java and Python, respectively.
The Java service in \autoref{code:example-java} exposes an endpoint \cc{/updateProfile} that accepts a role update request (line 2).
The service enforces user authentication via the \cc{@PreAuthorize()} annotation (line 1), verifying that the requester possesses valid credentials.
Upon receiving the request, the service retrieves the username and token from the current session (lines 4-5), constructs an HTTP request with authorization headers (lines 7-8), and forwards the request to the Python service at \cc{http://localhost:5000/setUserRole} (line 14).

The Python service in \autoref{code:example-python} exposes an endpoint \cc{/setUserRole} that receives the forwarded request (line 19).
The service extracts \cc{username} and \cc{role} from the request body and directly updates the user's role (lines 22-23).
Authentication is enforced at the router level, where the dependency \cc{Depends(authz)} (line 17) validates the JWT token's authenticity (lines 27-28).
However, the \cc{authz()} function is insufficient as it checks whether the user has the general permission to switch roles via \cc{can\_switch\_roles()}, but fails to validate whether the user is eligible for the specific role being requested, \eg, \cc{dev} or \cc{admin}.
This creates a severe privilege escalation vulnerability, allowing any authenticated user to assume higher-privileged roles.

\subsection{Challenges}
\label{s:challenges}
We revisit the three challenges introduced in \autoref{s:intro} using the code of the example.

\PP{C1: Distributed and Heterogeneous Architecture}
The vulnerability in \autoref{code:example} spans Java and Python services communicating via REST.
Detecting it requires identifying the inter-service call at line 14, tracing how the user-controlled \cc{role} parameter propagates across the service boundary to \cc{http://localhost:5000/setUserRole}, and verifying authorization in either service.
Traditional static analysis tools cannot perform such cross-language data flow analysis~\cite{mscan, mace, pcfinder}.

\PP{C2: Semantic Understanding}
Recognizing that \cc{setUserRole()} is a privileged operation requires understanding the function's security-critical purpose.
Moreover, determining whether the service-specific \cc{authz()} provides adequate protection requires reasoning about what \cc{can\_switch\_roles()} actually verifies—whether it checks general role-switching permission or validates eligibility for the specific role requested.
This semantic gap cannot be captured by manually enumerated patterns.

\PP{C3: Correlating Checks and Operations}
Determining which checks protect \cc{setUserRole()} in \autoref{code:example-python} requires correlating the operation with security checks through control and data dependencies.
From the endpoint decorator (line 19), the analysis must trace to the router definition (line 17), discover the \cc{Depends(authz)} dependency, and retrieve its implementation (lines 25-33) to establish the correlation.
Different applications structure these relationships differently, requiring the analysis to adapt rather than follow a predetermined pipeline.

\begin{table*}[t]
\centering
\caption{Fundamental query categories and their frequencies across 290 CodeQL queries.}
\label{tab:operation-frequency}
\small
\resizebox{\textwidth}{!}
{
\begin{tabular}{llllr}
\toprule
Symbol & Category & Description & Example & Frequency (\%) \\
\midrule
$Q_{name}$ & Name-based lookup & Match code elements using identifier names or keywords & Find \cc{update\_role()} & 286 (98.6\%) \\
$Q_{ast}$ & AST-based lookup & Identify code patterns by analyzing AST structures & Find all method calls & 214 (73.8\%) \\
$Q_{flow}$ & Flow tracking & Track how values propagate between program elements & Trace \cc{request} to sink & 205 (70.7\%) \\
$Q_{cg}$ & Call-graph traversal & Explore relationships between functions through the call graph & Find callers of \cc{authz()} & 144 (49.7\%) \\
\bottomrule
\end{tabular}
}
\end{table*}

\section{Design}
\label{s:design}
In this section, we present the design of \sys, our agentic program analysis framework for detecting privilege escalation vulnerabilities in polyglot microservices.
\sys addresses the challenges through the orchestrated interplay of LLM reasoning and static program analysis.

The overall architecture of \sys is illustrated in \autoref{fig:workflow}.
To detect privilege escalation vulnerabilities, we decompose the analysis into three interconnected components.
First, \sys iteratively identifies privileged operations by analyzing function names, documentation, comments, and source code to determine which operations require authorization checks.
Second, \sys conducts holistic cross-service data-flow analysis to trace how data propagates from external user inputs to the identified privileged operations.
Since attackers can manipulate user-controlled inputs, this analysis identifies potential attack vectors where malicious requests could reach privileged operations.
Third, \sys traverses the identified flows to locate authN/authZ checks along these paths.
When needed, \sys dynamically retrieves additional context to validate whether the checks verify appropriate properties (\eg, role eligibility vs. mere authentication).
\sys also checks the feasibility of the paths using a lightweight path constraint collection and satisfiability checking.

From a high level, \sys takes as input the source code of a microservice application and a prompt that describes the detection task.
The output is a set of privilege escalation vulnerability reports.
To enable this workflow, we design a code search engine (\autoref{s:design-search-engine}) that provides the building blocks for \sys to navigate codebases and perform program analysis.
Next, we describe privileged operation identification in \autoref{s:design-privilege}, explain cross-service flow tracking in \autoref{s:design-flow}, and present security check validation in \autoref{s:design-validation}.

\subsection{Code Search Engine}
\label{s:design-search-engine}
We frame static program analysis as a code search problem over the target application's codebase, where vulnerability detection logic is expressed as code queries~\cite{joern, codeql}.
Formally, given an application $A$, a code search query $Q(A)$ returns matching code components when the query is executed.
Complex program analysis can be composed by chaining multiple searches, where subsequent queries operate on the results of preceding queries.
An LLM agent can dynamically plan and adapt its search strategy by composing these code search primitives based on intermediate results.

Prior query-based solutions such as Joern~\cite{joern} and CodeQL~\cite{codeql} have been highly successful and widely adopted in practice.
They provide expressive domain-specific query languages that offer human developers fine-grained control and flexibility to specify complex security queries.
However, this expressiveness comes with complexity that hinders its use by LLM agents.
The large syntax space and extensive API surface (\eg, several thousand APIs in Joern and CodeQL) make these query languages challenging for LLMs to learn and generate correctly, especially given their limited resources in LLM pretraining data.
Moreover, these tools require different query syntax for the same semantic concept across programming languages.
For instance, querying method calls in CodeQL requires \cc{MethodAccess} for Java but \cc{Call} for Python, each with different properties and predicates to access callers, arguments, and return values.
This complexity leads to frequent errors when LLMs attempt to generate queries, such as using incorrect syntax or mixing constructs across languages.
In an agentic design that requires multiple rounds of iterative querying, such errors become particularly costly, as each mistake requires a great number of additional LLM calls to correct and retry (see \autoref{s:eval-ablation} for details).

To address these challenges, we design the code search engine with a small set of \emph{code search primitives} to enable scalable and flexible code search (\autoref{s:search-query}).
We also present a set of property functions for obtaining detailed information from code components (\autoref{s:search-query-property}).

\subsubsection{Code Search Primitives}
\label{s:search-query}
To identify the core query operations needed for static security analysis, we examined how existing CodeQL queries detect vulnerabilities.  
Specifically, we analyzed 290 security detection queries in the ``Security'' folder of CodeQL's query library (version~2.22.4)~\cite{codeqlqueries} for C++ and Java.  
Through manual inspection, we identified four fundamental query operations.
These primitives represent different levels of reasoning over program structures, from simple text matching to interprocedural analysis.
\autoref{tab:operation-frequency} summarizes their usage frequencies across the analyzed queries.
Name-based lookup (\(Q_{name}\)) matches code elements by identifier names or keywords to locate functions, variables, or types (98.6\%).
AST-based lookup (\(Q_{ast}\)) captures structural patterns in the program's abstract syntax tree (73.8\%).
Flow tracking (\(Q_{flow}\)) traces how values propagate between sources and sinks (70.7\%).
Call-graph traversal (\(Q_{cg}\)) analyzes interprocedural relationships, such as identifying callers or callees of specific functions (49.7\%).

These observations reveal that, despite differences in language and vulnerability type, most analyses rely on a small set of recurring query patterns. 
We therefore design \sys's code search engine around these four unified \emph{code search primitives}.
Each primitive abstracts away language-specific syntax and is exposed through a \emph{unified API} that LLM agents can invoke consistently across programming languages.
Unlike prior systems such as CodeQL, which require thousands of language-specific predicates and class hierarchies, our design provides a concise, semantically aligned interface.
This simplification allows LLMs to compose complex analyses through simple, chainable queries without memorizing tool-specific syntax.
By offering a small yet expressive set of primitives, \sys reduces query-generation errors while preserving the analytical depth required for comprehensive vulnerability detection.

\PP{Name-based Lookup $Q_{name}(service, name)$}
This query locates program elements (functions, classes, variables, \etc) in a service by their identifier names.
It is the most direct means of identifying security-critical code elements and retrieving on-demand context.
The operation supports both exact matching and regular expression patterns for fuzzy matching.
For example, $Q_{name}(service, ``update\_role")$ locates the exact operation \cc{update\_role()} that modifies user roles, while $Q_{name}(service, ``update.{*}")$ finds all functions whose names start with ``update".
This operation is also useful for identifying framework-specific handlers. For instance, $Q_{name}(service, ``{.}{*}Router")$ locates FastAPI router definitions, and $Q_{name}(service, ``auth.{*}")$ finds authorization-related functions like \cc{authz()}.

\PP{AST-based Lookup $Q_{ast}(service, operation)$}
This query identifies code locations that perform specific syntactic operations, such as field accesses, binary operations, or variable assignments.
We define a unified vocabulary of operation types that cover common code patterns across languages.
For example, $Q_{ast}(service, ``call")$ locates all function/method invocations, regardless of whether they are represented as \cc{MethodAccess} in Java, \cc{Call} in Python, or \cc{CallExpr} in JavaScript.
Similarly, $Q_{ast}(service, ``field\_access")$ finds all field or attribute accesses (\eg, \cc{obj.field}).
By providing a language-agnostic interface for common syntactic patterns, this operation enables the LLM agent to search for specific code constructs without learning the different AST node types in each language's analysis tooling.

\PP{Flow Tracking $Q_{flow}(service, from, to)$}
This query traces how data propagates from one program element to another within a service.
For example, $Q_{flow}(service, ``request", ``update\_role")$ traces how the user-provided \cc{role} parameter flows in \autoref{code:example-python}.
The operation returns the complete path at variable-level granularity, showing data propagation as \cc{request} $\rightarrow$ \cc{data.get("role")} $\rightarrow$ \cc{update\_role(username, role)}.
This enables the agent to identify that user-controlled data reaches a privileged operation.

\PP{Call Graph Traversal $Q_{cg}(service, function, direction)$}
This query provides bidirectional call graph navigation in a service, finding all callers of a given function or all callees it invokes.
For example, $Q_{cg}(service, ``update\_role", ``callers")$ identifies all code paths leading to the privileged operation, while $Q_{cg}(service, ``setUserRole", ``callees")$ reveals what operations a handler performs.
The agent can iteratively traverse the call graph to discover execution paths and verify security checks are present.
Note that precisely identifying call targets, especially for dynamic calls, remains an open challenge in static analysis.
Our implementation relies on CodeQL's call resolution, which provides sound approximations but may be conservative in complex scenarios involving reflection or dynamic dispatch.

\subsubsection{Property Functions}
\label{s:search-query-property}
For elements returned by query operations, we define a set of property functions that provide rich information to enable the LLM agent to reason about the elements.
Three property functions are particularly important.

\PP{getLocation()}
This function returns the element's location in the codebase, including file path, line number, and column position.
This allows the agent to precisely reference code locations when reporting vulnerabilities or retrieving surrounding context.
For example, after identifying a privileged operation, the agent can use \cc{getLocation()} to determine which service and file it resides in.

\PP{getSource()}
This function returns the source code of the element, ranging from a single statement to an entire function body.
This enables the agent to examine the actual implementation without additional queries.
For instance, after locating \cc{update\_role()} via $Q_{name}$, the agent can call \cc{getSource()} to analyze its implementation and determine whether it performs authorization checks.

\PP{getType()}
This function returns the type of the element.
Specifically, it performs type inference for variables to determine their types.
This is helpful when filtering query results based on value types or for identifying sanitizations where type conversions occur.

\subsection{Finding Privileged Operations}
\label{s:design-privilege}

In this work, we define privileged operations as call sites that access sensitive resources (\eg, user-specific records, configuration settings), perform security-critical actions (\eg, database writes, system commands), or modify protected state (\eg, permissions, authentication tokens, user roles).
To identify privileged operations, we provide the privileged operation definition in the prompt, along with instructions for using the code search engine, and instruct the LLM to generate queries to search the codebase.
The agent iteratively queries the codebase, validates results via an LLM, and outputs the locations of privileged operations.
Optionally, \sys analyzes application documentation (\eg, README files) to understand the application's purpose and generate more targeted queries.
For example, after understanding the motivating example is a user management system, \sys generates queries like $Q_{name}(service, ``update.*")$ to locate state modification operations and $Q_{name}(service, ``.*role.*")$ to find role-related functions.
These regular expression queries allow fuzzy matching, discovering operations like \cc{update\_role()}, \cc{setUserRole()}, and \cc{assign\_role()}.
\sys can also use $Q_{ast}$ queries to find specific operation types, such as $Q_{ast}(service, ``method")$ to locate all methods (functions) defined in the service.

Because queries may match many sites beyond actual privileged operations, \sys must validate each candidate.
For each candidate, \sys retrieves the source code using \cc{getSource()} and then consults the LLM to assess whether this is a privileged operation, classifying it accordingly based on the provided definitions.
For instance, when validating \cc{update\_role()}, \sys observes that it modifies user roles in the database, a security-critical action that changes user privileges.
The process continues with additional query rounds if the iteration limit has not been reached, allowing \sys to refine its search based on discovered patterns.

\subsection{Identifying Cross-service Flows}
\label{s:design-flow}
After identifying privileged operations, \sys must determine which operations are reachable from external user inputs (\ie, there is a data flow path), as only reachable operations can be exploited by attackers for privilege escalation.
The code search primitives described earlier operate within individual services.
Privilege escalation vulnerabilities in microservices often span multiple services, requiring analysis to trace data flow across service boundaries.
While the agent could in principle chain the primitives to perform cross-service analysis, this would require complex multi-step reasoning to identify service entry points, match inter-service calls to their target endpoints, and stitch together flows across service boundaries.
Such a process is often prone to errors and inefficiency.
We thus define three additional operations to facilitate cross-service privilege escalation detection.

\PP{Source Identification $Q_{source}(service)$}
This query identifies data sources in a service, \ie, locations where untrusted data from external sources enters the service (\eg, HTTP request parameters, message queue consumers, API endpoints).
We also define $Q_{user}$---a special case of $Q_{source}$, which identifies external user inputs entering the entire system, typically at the proxy or gateway service.
We reused the method in MScan~\cite{mscan} to identify user sources by prompting the LLM to analyze the gateway configuration file.

\PP{Inter-service Communication $Q_{inter}(service)$}
This query identifies inter-service communication points, including outgoing HTTP calls, RPC invocations, Kafka messages, WebSocket connections, GraphQL queries, \etc
These represent points where data flows from one service to another.
Each inter-service call is identified by a unique channel identifier that associates the caller and callee.
This design is inspired by prior work~\cite{mscan}.
Consider the example in \autoref{code:example}.
The outgoing HTTP call in the Java service uses the URL \cc{http://localhost:5000/setUserRole} as its channel identifier, which \sys matches to the corresponding endpoint \cc{/setUserRole} in the Python service.
This enables \sys to connect flows across the language/service boundary.

\PP{Global Flow $Q_{globalflow}(source, sink)$}
This query traces data flow across multiple services from a source element to a sink element, connecting intra-service flows through inter-service communication points.
Unlike $Q_{flow}$ that operates within a single service, $Q_{globalflow}$ constructs a global reachability graph.

\begin{figure}[t]
    \center{\includegraphics[width=0.95\columnwidth]{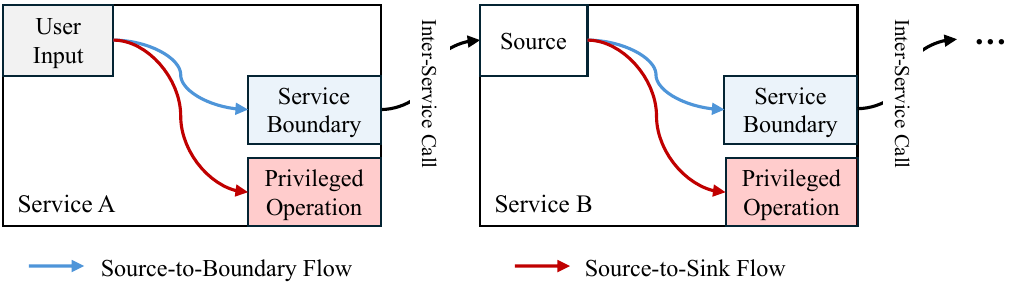}}
    \caption{Illustration of inter-service analysis.}
    \label{fig:globalflow}
\end{figure}

As illustrated in \autoref{fig:globalflow}, a data source entering a service can flow through two types of paths:
(1) a source-to-boundary flow that reaches the service boundary to another service via an inter-service call, and (2) a source-to-sink flow that directly reaches a privileged operation.
Inter-service calls act as bridges, connecting flows across service boundaries.
$Q_{globalflow}$ operates in two phases.
It first performs intra-service flow analysis using $Q_{flow}$ to track these paths, and then chains them into a global inter-service view by matching inter-service calls identified by $Q_{inter}$ to their target endpoints identified by $Q_{source}$.

\autoref{alg:cross-service-flow} presents the algorithm implementing $Q_{globalflow}$, which takes as input the set of services $\mathcal{S}$ and identified privileged operations $\mathcal{P}$, and outputs a global reachability graph $\mathcal{G}$.
In the first phase, for each service $s$, the algorithm identifies user input entry points using $Q_{source}(s)$ (line 4), privileged operations $\mathcal{P}$ within the service (line 5), and outgoing inter-service calls using $Q_{inter}(s)$ (line 6).
For each user input source, the algorithm uses $Q_{flow}(s, src, dst)$ to check if data can flow to privileged operations or inter-service calls (line 10).
If a flow exists, an edge is added from source to sink in the global graph $\mathcal{G}$.
In the second phase, the algorithm connects flows across service boundaries by linking each inter-service call to its target endpoint in the receiving service (lines 16-17).
This leverages the approach from MScan~\cite{mscan}, which assigns each inter-service communication channel a unique identifier to correlate outbound calls in one service with inbound data in the other service.
Finally, $Q_{globalflow}$ returns all paths from external user inputs ($Q_{user}$) to privileged operations $\mathcal{P}$.
These flows represent potential authorization vulnerabilities where user-controlled data can reach security-critical operations across service boundaries.

With these operations, \sys can identify global cross-service flows that can reach the privileged operations.

\SetAlCapNameFnt{\small}
\SetAlCapFnt{\small}
\SetAlFnt{\footnotesize}
\begin{algorithm}[t]
    \caption{Inter-service flow analysis.}
    \footnotesize
    \label{alg:cross-service-flow}
    \SetKwInOut{Input}{Input}
    \SetKwInOut{Output}{Output}
    \SetKwFunction{QFlow}{$Q_{flow}$}
    \SetKwFunction{QInter}{$Q_{inter}$}
    \SetKwFunction{QSrc}{$Q_{source}$}
    \Input{Services $\mathcal{S}$, Privileged operations $\mathcal{P}$}
    \Output{Global reachability graph $\mathcal{G}$}
    
    $\mathcal{G} \leftarrow \emptyset$\\
    
    \algcomment{Phase 1: Intra-service flow analysis}\\
    \ForEach{$s \in \mathcal{S}$}
    {
        $Srcs \leftarrow$ \QSrc($s$)\\
        $PrivOps \leftarrow \{p \in \mathcal{P} \mid p \in s\}$\\
        $InterCalls \leftarrow$ \QInter($s$)\\
        
        \algcomment{Check reachability and add edges}\\
        \ForEach{$src \in Srcs$}
        {
            \ForEach{$dst \in PrivOps \cup InterCalls$}
            {
                \lIf{$\QFlow(s, src, dst) \neq \emptyset$}{
                    $\mathcal{G}$.addEdge($src$, $dst$) \algcomment{Add path to G} 
                }
            }
        }
    }
    
   \algcomment{Phase 2: Inter-service connection}\\
    \ForEach{inter-service call $c: s_1 \to s_2$ in $\mathcal{G}$}
    {
        $endpoint \leftarrow c$.targetEndpoint in $s_2$\\
        $\mathcal{G}$.addEdge($c$, $endpoint$) \algcomment{Cross-service edge}\\
    }
    
    \Return $\mathcal{G}$
\end{algorithm}

\subsection{Validating Flows}
\label{s:design-validation}
\sys validates whether appropriate authN/authZ checks exist along these flows and whether the paths are feasible. 
This validation leverages the LLM's reasoning capabilities combined with \emph{on-demand context retrieval} to handle diverse authN/authZ patterns across different frameworks and implementation styles.
The agent traverses flow paths node by node (\eg, function calls, conditionals), deciding at each point whether current information suffices or whether additional context is needed.
At any point, the agent can flexibly issue code search primitives or property functions to retrieve specific code snippets across the entire codebase.
Unlike prior code agents~\cite{openhands, sweagent, enigma} that directly load entire source files into the LLM context, \sys reads only the structured results returned by search operations—such as function signatures, call relationships, or targeted code snippets.
This abstraction achieves \emph{scalability} and \emph{efficiency} by avoiding context window exhaustion and enabling cross-language navigation through language-agnostic search results.

\PP{AuthN/AuthZ Check Localization}
\sys traverses each flow to locate authN/authZ checks by analyzing decorators, middleware, or inline validation logic.
For example, in \autoref{code:example-python}, when encountering \cc{@router.post(``/setUserRole")} (line 19) without explicit authorization, \sys queries $Q_{flow}(service, \_, ``@router")$ to locate the router definition (line 17) and discovers the \cc{Depends(authz)} dependency.
To understand what this dependency does, \sys retrieves \cc{authz}'s implementation (lines 25-33) via $Q_{name}$ and \cc{getSource()}, confirming it validates JWT tokens.
This iterative retrieval enables \sys to discover authentication mechanisms regardless of their locations in the code structure.

\PP{Sufficiency Assessment}
Beyond locating checks, \sys must determine whether they adequately protect the privileged operation—a fundamentally semantic reasoning task.
The agent performs this analysis by consulting the LLM with both the check's implementation and the privileged operation's context.
The agent first asks the LLM to classify the check as authN or authZ, and if authZ, to identify its specific type (\eg, role-based, permission-based, resource ownership).
The agent then examines what protection the privileged operation requires and identifies any semantic gap between what is checked and what is required.
For instance, the insufficient authZ in \autoref{code:example-python} could be correctly labeled by leveraging LLM's semantic understanding of security properties.

\PP{Path Constraint Validation}
Static analysis typically cannot determine whether a flow is actually feasible or reachable, resulting in false positives.
\sys mitigates this with a conservative path constraint validation stage to filter out infeasible paths.
Specifically, \sys leverages the LLM to extract and collect path constraints from the identified flows, and prunes flows with unsatisfiable constraints.
For each flow, the LLM traverses intermediate nodes in the flow and identifies conditional guards that constrain the path.
The LLM then translates these constraints into logical predicates in SMT-LIB format, which are then sent to the Z3 solver~\cite{z3solver} for satisfiability checking.
If the Z3 solver reports unsatisfiable constraints, \sys marks it as a false positive.

Since the path constraints are collected by the LLM, they may introduce inaccuracies. 
For example, the generated constraints might be syntactically invalid or simply incomplete.
We thus instruct the LLM to be \emph{conservative}.
If the LLM determines that the code logic is too complex to be reliably modeled as a constraint (\eg, multiple layers of data dependencies), it skips this step and retains the uncertain cases as potential vulnerabilities.

\section{Implementation}
\label{s:impl}

We implemented a prototype of \sys using 5.8K lines of CodeQL queries and 4.3K lines of Python code.
\sys supports the \numlang most popular programming languages for cloud applications: Go, Java, JavaScript (JS), Python, C\#, C, and C++.
We make \sys and the prompts available at \url{https://github.com/columbia/neo}.
We next present several important implementation details.

\PP{Agent Orchestration}
\sys implements the agentic loop in Python, coordinating LLM reasoning with program analysis primitives.
The agent has access to a set of utility tools, including privileged operation search, data-flow analysis, call graph traversal, context retrieval, and validation tools.
Each tool is realized either through carefully designed prompts that consult the LLM or through CodeQL queries.
At each iteration, the LLM receives the current analysis state and generates the next action by executing a search primitive, requesting context, or performing validation.
The agent stops when no new privileged operations are found, all flows are validated, or an iteration limit is reached.

\sys's task prompt describes the three-step workflow (\autoref{fig:workflow}) and exposes the search primitives as well as the fallback option of bash commands to the LLM.
We incorporate the application’s documentation to allow the agent to tailor its analysis to each application.
Two true vulnerabilities are also provided as demonstrations.

\PP{Code Database Construction}
\sys builds upon CodeQL's infrastructure to represent the target application's source code in a structured database.
The database captures multiple program representations, including abstract syntax trees for syntactic structure, control-flow graphs for execution paths, data-flow graphs for tracking data dependencies, and call graphs for inter-procedural relationships.
For compiled languages like Java, the extraction analyzes compilation artifacts and intermediate representations.
For interpreted languages like Python, the extraction directly parses source code to construct the database.
Database construction is performed once per application.

\PP{Communication Channel Identification}
We implemented $Q_{inter}$ by developing comprehensive, language-specific CodeQL queries to identify communication patterns on both the caller and callee sides.
To extract the channel identifier for each cross-service communication point, our implementation performs backward data-flow analysis from the communication call site to identify the string literal or constant defining the channel (\eg, URL, topic name).
Our implementation currently supports major protocols including REST, gRPC stubs, Kafka, RabbitMQ, WebSocket, and Dubbo.

\PP{CodeQL-based Search Primitives}
We implemented the code search primitives as CodeQL templates with a few placeholders and developed Python scripts to dynamically fill them.
When \sys issues commands like $Q_{flow}(source, sink)$, \sys dynamically populates these templates with the specific method names or types identified by the LLM.
Each primitive is implemented as a separate CodeQL module that abstracts language-specific CodeQL predicates into a unified interface.
For example, the $Q_{ast}$ primitive for method calls translates to \cc{MethodAccess} in Java, \cc{Call} in Python, and \cc{CallExpr} in JavaScript, but presents a single API to the agent.
This design ensures that the LLM agent can query codebases without knowledge of language-specific analysis constructs.

\PP{SMT-LIB Constraint Generation}
To generate the constraints for path validation, we provide the identified flows to the LLM and instruct it to convert the conditional statements (\eg, \cc{if}, \cc{switch}) directly into SMT-LIB format.
Specifically, similar to recent work~\cite{llmsa, llift, concollmic}, the LLM is prompted to declare relevant variables, define constants for status codes or fixed values, and assert the logical predicates encountered along the path. 
This process relies on the LLM’s semantic understanding to interpret diverse coding patterns (\eg, string comparisons or status flag checks) that are often difficult for traditional symbolic execution to model across different languages.

\section{Evaluation}
\label{s:eval}

We extensively evaluate \sys to answer the following research questions:
\squishlist
\item \PP{Effectiveness}
Can \sys detect previously unknown vulnerabilities in real-world microservice applications?

\item \PP{Ablation Study}
What is the contribution of each component of \sys to its overall effectiveness?

\item \PP{Comparison with Prior Work}
How does \sys perform compared to state-of-the-art analysis tools?

\item \PP{Efficiency and Cost}
What are the runtime and LLM costs of running \sys?

\item \PP{Extensibility}
How applicable is \sys to other vulnerability analysis tasks?
\squishend

\begin{table}[t]
\centering
\caption{Testing dataset of \numtestapp microservice applications.
\cc{\# Stars} denotes the number of GitHub stars.
}
\renewcommand{\arraystretch}{1.25}
\label{tab:evaluation-dataset}
\resizebox{\columnwidth}{!}
{
\begin{tabular}{lrrl} 
\toprule
App & \# Stars & \# LoC & Languages \\
\midrule
\rowcolor{lightgray} light-reading-cloud~\cite{lightreadingcloud} & \num{1465} & \num{6394} & Java \\
PiggyMetrics~\cite{piggymetrics} & \num{13757} & \num{19942} & JS, Java \\
\rowcolor{lightgray} Pitstop~\cite{pitstop} & \num{1145} & \num{96576} & JS, C\# \\
SiteWhere~\cite{sitewhere} & \num{1034} & \num{40764} & Java  \\
\rowcolor{lightgray} Supermarket~\cite{supermarket} & \num{2087} & \num{167169} & Java  \\
Food Delivery~\cite{fooddeliverydotnet} & \num{941} & \num{178063} & JS, C\# \\
\rowcolor{lightgray} Online Boutique~\cite{onlineboutique} & \num{19254} & \num{31179} & JS, Java, Go, Python, C\# \\
Booking~\cite{bookingmicroservices} & \num{1256} & \num{34974} & JS, C\# \\
\rowcolor{lightgray} Hotel Map~\cite{gohotelmap} & \num{1085} & \num{2696} & Go \\
JBone~\cite{jbone} & \num{1008} & \num{12360} & Java \\
\rowcolor{lightgray} Train Ticket~\cite{trainticket} & \num{836} & \num{344119} & JS, Java, Go, Python \\
Mall4cloud~\cite{mall4cloud} & \num{5933} & \num{116043} & JS, Java \\
\rowcolor{lightgray} AspnetRun E-Shop~\cite{aspnetruneshop} & \num{3121} & \num{62105} & C\# \\
Cinema~\cite{cinemamicroservice} & \num{1773} & \num{8152} & JS \\
\rowcolor{lightgray} TODO App~\cite{polyglottodo} & \num{1409} & \num{16240} & JS, Java, Go, Python \\
ABP eShopOnAbp~\cite{abp-eshop} & \num{736} & \num{220134} & JS, C\# \\
\rowcolor{lightgray} Spring Boot Basics~\cite{springbootbasics} & \num{728} & \num{5046} & JS, Java \\
Magda~\cite{magda} & \num{564} & \num{625045} & JS, Java \\
\rowcolor{lightgray} Genie~\cite{genie} & \num{1756} & \num{127889} & JS, Java, Python\\
DeathStarBench~\cite{DeathStarBench} & \num{871} & \num{3621236} & JS, C, C++, Go, Python\\
\rowcolor{lightgray} Swarm~\cite{mallswarm} & \num{12673} & \num{133793} & Java \\
\bottomrule
\end{tabular}
}
\end{table}

\subsection{Experimental Setup}
\PP{Dataset}
To comprehensively evaluate \sys's performance, we use two datasets: (1) an evaluation corpus of \numtestapp open-source microservice applications, and (2) a ground-truth dataset containing 20 verified privilege escalation vulnerabilities across \numgroundapp applications.

\emph{Evaluation corpus.}
We constructed an evaluation corpus of \numtestapp microservice applications by searching open-source repositories from GitHub with the keyword ``microservice''.
Since several thousand repositories were returned in the search,  we applied multiple filtering strategies to ensure quality and representativeness.
First, each application must have at least \num{500} stars, indicating significant community adoption and popularity.
Second, applications must be actively maintained, defined as having commits or releases within the past three years, to reflect current development practices.
We also prioritized applications used in prior microservice research~\cite{mscan, mspolicy} to enable direct comparison with existing work.
\autoref{tab:evaluation-dataset} presents the dataset details, including lines of code (LoC) for each application.
These applications span a wide range of complexity, from systems with a few services to large-scale benchmarks with dozens of services.
They exhibit significant diversity in both application domains (\eg, e-commerce, ticketing) and programming languages (\eg, Java, Go, Python, C\#), with most being polyglot systems.

\emph{Ground-truth dataset.}
We constructed a ground-truth dataset of applications with verified privilege escalation vulnerabilities to assess false negatives.
Specifically, we searched MITRE CVE~\cite{cve-org} and NVD~\cite{nvd} using the application names and identified previously reported privilege escalation vulnerabilities in these applications.
The dataset includes \numgroundappvul previously reported vulnerabilities across \numgroundapp applications, as well as \numnewvulbench previously unreported vulnerabilities uncovered by \sys, yielding 20 verified vulnerabilities in total.
Details are shown in \autoref{tab:dataset-ground-truth}.

\PP{Evaluation Procedures}
To run \sys, we first compile each application into a code database on which we can apply code search primitives.
We then launch \sys and instruct it to autonomously detect privilege escalation vulnerabilities with a time budget of 10 hours per application.
By default, we use Claude Sonnet 3.7 for agent reasoning with a temperature of 0.2.
As for the in-context examples, we selected two vulnerabilities from applications outside the evaluation dataset to ensure the model generalizes its security reasoning.
These examples were used across all experiments to provide consistent guidance.
The experiments were conducted on a CPU server with two AMD EPYC 7502 32-Core processors and 251GB of RAM running Ubuntu 24.04.3 LTS.

\begin{table}[t]
\centering
\scriptsize
\caption{Ground-truth dataset with 20 verified vulnerabilities across \numgroundapp applications, including \numgroundappvul previously reported vulnerabilities and \numnewvulbench previously unreported vulnerabilities uncovered by \sys in Newbee Mall during evaluation.}

\label{tab:dataset-ground-truth}
\resizebox{\columnwidth}{!}
{
\begin{tabular}{lrrlc} 
\toprule
App & \# Stars & \# LoC & Languages &\# Vuls.\\
\midrule
Newbee Mall~\cite{newbeemall} & \num{11437} & \num{172057} & Java & 11 + 2\\
ZLT Platform~\cite{zltplatform} & \num{4721} & \num{36950} & Java, JS & 3\\
Armeria~\cite{armeria} & \num{5033} & \num{633902} & Java, JS & 2\\
Spring-cloud-dataflow \cite{springclouddataflow} & \num{1138} & \num{235572} & Java, JS, Python & 2 \\
\bottomrule
\end{tabular} 
}
\end{table}

\subsection{Effectiveness}
\label{s:eval-effectiveness}

We first evaluated the effectiveness of \sys in detecting privilege escalation vulnerabilities and summarized the detection results in \autoref{tab:effectiveness}.
In total, \sys identified 39 true-positive vulnerabilities out of 51 reports across the two datasets.
\numnewvul out of the 39 are new privilege escalation vulnerabilities.
Specifically, in the evaluation corpus, \sys uncovered \numnewvulcorpus previously unknown vulnerabilities.
In the ground-truth dataset, \sys found 17 vulnerabilities in total, including \numnewvulbench previously unreported vulnerabilities; this results in a recall of 85.0\% with a precision of 81.0\%.
For all the reported cases from \sys in both datasets, we verified the vulnerabilities by manually analyzing the vulnerable flows from application entry to the privileged operations.
We constructed proof-of-concept exploits to demonstrate the exploitability of these vulnerabilities.
Overall, there were a total of 12 false positives generated, achieving an overall precision ($\frac{TP}{TP+FP}$) of 76.5\%.

\PP{Characterization of True Positives}
During the verification process, we found that these vulnerabilities showcase diverse security impacts, mostly determined by the variety of privileged operations.
Standard ones include database operations and file operations, which can be exploited to perform SQL injection, arbitrary file reads and writes, or path traversal attacks.
We also identified 18 application-specific privileged operations.
These application-specific operations include privilege elevation regarding user roles or permissions, administrative command execution (\eg, system configuration changes), and critical business logic operations (\eg, payment processing or order fulfillment).
This diversity demonstrates that privilege escalation vulnerabilities in microservices extend beyond traditional attack vectors and require a comprehensive analysis of application-specific security contexts.
They are hard to (automatically) identify without \sys's LLM-based reasoning.
We further present a case study in \autoref{s:eval-case-study}.

\begin{table}[t]
\centering
\scriptsize
\caption{Privilege escalation detection results. In the evaluation corpus, all true positives are previously unknown vulnerabilities. In the ground-truth dataset, \sys reports 17 true positives in total, including \numnewvulbench previously unreported vulnerabilities.}
\label{tab:effectiveness}
{
\begin{tabular}{lccccc} 
\toprule
Dataset & TP & FP & FN & Prec. & Recall\\
\midrule
Eval. corpus & \numnewvulcorpus & 8 & N/A  & 73.3\%& N/A\\
Ground-truth dataset & 17 & 4 & 3 & 81.0\% & 85.0\%\\
\midrule
Total & 39 & 12 & N/A & 76.5\% & N/A\\
\bottomrule
\end{tabular} 
}
\end{table}

\PP{False Positives}
We systematically analyzed the 12 false positives to understand their root causes and identified two main categories.
First, 8 cases resulted from misidentifying the execution context, leading to over-approximation of security impacts.
Specifically, \sys incorrectly classified client-side operations as server-side vulnerabilities.
User-facing microservice applications often contain components that execute in the user's browser, such as browsing local directories for file uploads or displaying user-controlled content.
While \sys correctly identified that user input influenced these operations, they actually execute client-side and are legitimately controlled by users themselves.
The security impact is confined to the user's local environment with no server-side privilege escalation risk.
This confusion arose because \sys failed to distinguish between client-side and server-side execution contexts.
Second, the remaining 4 false positives resulted from \sys failing to detect authN/authZ checks.
While \sys correctly identified data flows to privileged operations, these operations were actually protected by authN/authZ mechanisms that \sys did not locate.
For example, checks implemented through external services or configuration files were not visible in \sys's analysis.

\PP{False Negatives}
We analyzed the 3 false negatives in the ground-truth dataset to understand \sys's limitations.
One case was caused by \sys failing to detect the privileged operation that was invoked through uncommon API patterns.
The second case was caused by a missed flow, even though the privileged operation was present.
The third case reveals a limitation in handling framework-specific URL processing semantics.
Specifically, CVE-2023-38493~\cite{CVE-2023-38493} exploited matrix variables---a URL feature that allows embedding key-value pairs within path segments using semicolons (\eg, \cc{/path;key=value/resource}).
An attacker could bypass authentication by sending \cc{/important;a=b/resources} to access a protected endpoint configured with \cc{decoratorUnder("/important/", authService)}.
While \sys correctly identified the authentication decorator as a security check, it failed to recognize that Spring's routing logic would normalize the matrix variable path to \cc{/important/resources}, causing the request to bypass the decorator's path matching.
This occurred because \sys failed to recognize this framework-specific URL parsing that could alter access control decisions at runtime.

\summarybox{
\sys detected \numnewvul new zero-day privilege escalation vulnerabilities in real-world microservices, and achieved 85.0\% recall and 81.0\% precision on the ground-truth dataset.
}

\subsection{Ablation Study}
\label{s:eval-ablation}

We conduct an ablation study to evaluate the performance of individual components in \sys and how they contribute to the overall performance.
We also characterize different LLM variations.

\subsubsection{Detection Effectiveness}
We design the following variants of \sys to understand the vulnerability detection effectiveness of \sys's components. 
We evaluate these variants on 42 vulnerabilities described in \autoref{s:eval-effectiveness}, including 22 vulnerabilities from the evaluation corpus and 20 vulnerabilities from the ground-truth dataset.

\PP{\sys-basic-sink}
A major component of \sys is identifying privileged operations beyond common patterns.
We thus design \sys-basic-sink to use only a list of standard privileged operations (\eg, networking, database, and file access) as adopted in prior work~\cite{mscan, mace} without application-specific privileged operations generated by LLMs.

\PP{\sys-no-queryop} 
\sys employs code search primitives for flexible code search. 
In \sys-no-queryop, we remove these query operations.
Instead, we require \sys-no-queryop to directly generate CodeQL queries for code search.

\PP{\sys-no-odctx}
On-demand context retrieval is another key technique in \sys.
To understand its necessity, \sys-no-odctx allows only a one-time context retrieval after the cross-service flow is produced, instead of doing it iteratively.

\begin{table}[t]
\centering
\scriptsize
\caption{Detection effectiveness in the ablation study. Recall is computed with respect to the total of 42 vulnerabilities (\numnewvul new and \numgroundappvul known) described in \autoref{s:eval-effectiveness}.}  
\label{tab:ablation}
{
\begin{tabular}{lcccrr}
\toprule
Variants & TP & FP & FN & Prec. & Recall\\
\midrule
$\sys_{basic-sink}$ & 14 & 0 & 28 & 100.0\% & 33.3\% \\
$\sys_{no-queryop}$ & 2 & 0 & 40 & 100.0\% & 4.8\% \\
$\sys_{no-odctx}$ & 27 & 31 & 15 & 46.6\% & 64.3\% \\
\sys & 39 & 12 & 3 & 76.5\% & 92.9\% \\
\bottomrule
\end{tabular} 
}
\end{table}

\PP{Results}
\autoref{tab:ablation} presents the results of our ablation study on the 42 vulnerabilities mentioned in \autoref{s:eval-effectiveness}.
Again, the full-fledged \sys system correctly identified 39 out of 42 vulnerabilities with only 3 false negatives.
In contrast, all three ablated variants show significantly degraded performance, demonstrating that each component plays a critical role in \sys's effectiveness.

\sys-basic-sink with only standard privileged operations identified 14 vulnerabilities (missed 28 vulnerabilities) with no false positives, achieving substantially lower recall of 33.3\%.
This result demonstrates that basic operations are insufficient for comprehensive vulnerability detection and cannot cover many privilege escalation vulnerabilities involving application-specific privileged operations.
This is also consistent with our characterizations of true positives mentioned in \autoref{s:eval-effectiveness}.

Removing the code search primitives in \sys-no-queryop resulted in the most dramatic performance degradation.
This variant detected only 2 vulnerabilities while missing 40, with a recall dropping to 4.8\%.
The results indicate that direct CodeQL query generation is extremely challenging, and LLMs mostly generate invalid CodeQL queries with a lot of syntax errors.
The code search primitives provide essential abstractions that enable \sys to systematically search for security-relevant code patterns.

\sys-no-odctx with only a one-time context retrieval identified 27 vulnerabilities, but produced 31 false positives and missed 15 vulnerabilities.
The substantial decline in both precision and recall underscores the importance of iterative context retrieval for handling complex vulnerability cases.
Without the ability to dynamically gather and integrate additional context during analysis, \sys struggles to locate authN/authZ checks, leading to both missed detections and incorrect assessments.

\subsubsection{Code Search Primitives}
Beyond vulnerability detection, we evaluate the correctness of the search results produced by the code search primitives. 
Specifically, we aim to determine whether search execution yields false positives (\ie, incorrect results) or false negatives (\ie, missing results). 
However, such evaluation is challenging, as it requires constructing a complete ground truth for the expected query results across all application code.
One potential approach is to use dynamic analysis to collect execution traces, but this method suffers from significant coverage limitations~\cite{lu2019does}. 
To bridge this gap, we conducted a rigorous manual assessment on a single application, PiggyMetrics~\cite{piggymetrics}, rather than on the entire dataset. We first recorded the complete trajectory of 132 unique executed code search primitives and their corresponding results during the analysis. From this set, we performed stratified random sampling, selecting 5 instances for each primitive type ($Q_{name}$, $Q_{ast}$, $Q_{flow}$, and $Q_{cg}$) to ensure a representative distribution across different search behaviors.
This resulted in a total of 20 searches for manual audit.
To establish a reliable ground truth, we exhaustively analyzed the target codebase for each search to verify the accuracy of the returned results and pinpoint the exact nature of any observed failures.

\PP{Imprecise Search Results}
Among the 20 sampled searches, we observed imprecise results in 7 instances, particularly within the types $Q_{flow}$ and $Q_{cg}$. These searches are not entirely incorrect but often return over-approximated results due to the inherent limitations of static analysis. For example, due to over-approximation in virtual dispatch, CodeQL's \texttt{polyCalls} (and similar predicates) frequently included edges to all potential method implementations because it could not statically determine the specific runtime target. These issues are potentially fixable by refining the underlying predicates in CodeQL, which we leave for future work. We encountered no imprecision in $Q_{name}$ and $Q_{ast}$.

\PP{Missing Search Results}
Among the 20 sampled searches, we observed that results were missing in a total of 5 instances, specifically within the $Q_{flow}$ and $Q_{cg}$ categories. These searches often return partial results due to the conservative nature of static analysis when handling complex control flow or implicit dependencies. Similar to the imprecise results, these omissions highlight the challenges of achieving perfect recall in automated code analysis.

\subsubsection{Validation Strategies}
\label{s:eval-ablation-validation}
\begin{figure}[t]
    \centering
        \centering
        \includegraphics[width=0.5\linewidth]{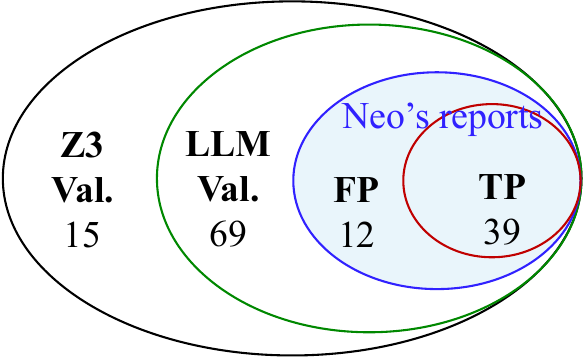}
        \caption{Impact of validation strategies. There are 135 flows initially.}
\label{fig:validation-impact}
\end{figure}

To understand how the two validation strategies help reduce false positives, we present the statistics in \autoref{fig:validation-impact}.
In our dataset, the cross-service flow analysis initially produced 135 flows from user inputs to privileged operations.
Among them, \sys generated constraints for 28 cases, all of which were in syntactically valid SMT-LIB format.
15 flows (11.1\%) were filtered out by proving through constraint solving that their corresponding constraints are infeasible.
We manually confirmed that all 15 cases were true negatives without any false negatives introduced.

Subsequently, the LLM-based authN/authZ validation examined the remaining flows and eliminated 69 that lack proper authN/authZ checks but are semantically benign based on code context.
Together, these two strategies reduced the initial 135 flows to 51 final reports, eliminating 84 potential false positives (a 62.2\% reduction).
This analysis highlights that both validation strategies help eliminate false positives.

To assess the precision of the LLM-based authN/authZ validation component, we manually audited 10 randomly sampled cases from the 69 flows.
By cross-referencing the LLM's justifications with the source code, we confirmed that in all 10 instances, the LLM correctly identified the explicit decorators (\eg, \cc{@RequiresAdmin()}) or implicit logic that protects the privileged operations.
This suggests that this validation strategy successfully reduces false positives without introducing false negatives in our samples.
However, we acknowledge that this LLM-based reasoning is inherently probabilistic and may occasionally yield incorrect classifications.
To mitigate this risk, one could adopt a human-in-the-loop approach~\cite{humanloop}, allowing security engineers to check the LLM's justifications.

\begin{figure*}[t]
    \centering
        \centering
        \includegraphics[width=0.98\linewidth]{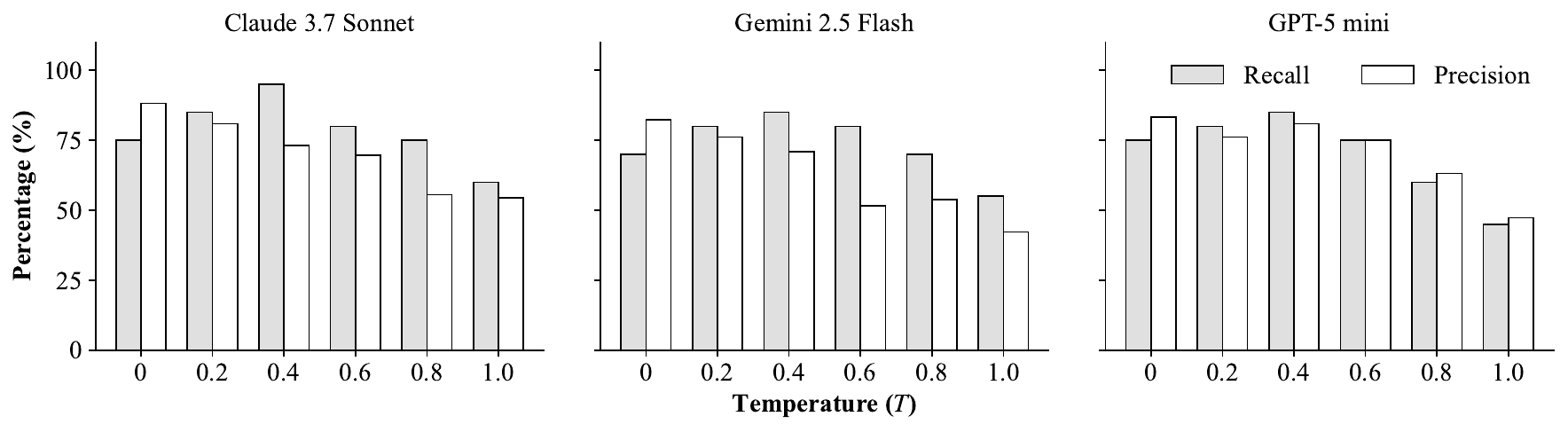}
\caption{Impact of model selection and temperature on vulnerability detection.}
\label{fig:ablation-model}
\end{figure*}

\subsubsection{LLM Variations}
\label{ss:llm-variations}

\sys relies on LLMs for core reasoning tasks. In this section, we analyze how performance is influenced by model selection, model temperature (stochasticity), and prompting strategy. 
Due to the high cost of model APIs (\autoref{s:eval-cost}), running this sensitivity analysis on the complete dataset is computationally and financially prohibitive.
Therefore, we evaluate this sensitivity analysis on the ground-truth dataset used in \autoref{s:eval-effectiveness}, which includes the previously reported vulnerabilities and the additional vulnerabilities uncovered by \sys in those same applications (\autoref{tab:effectiveness}).
Accordingly, recall in this subsection is measured over the ground-truth dataset.

\PP{Models and Temperatures}
We evaluated \sys with three different models: (1) Claude 3.7 Sonnet, which is the default model throughout our experiments, (2) Gemini 2.5 Flash from Google, and (3) GPT-5 mini from OpenAI. 
We analyzed the effect of stochasticity by testing temperatures $T \in \{0, 0.2, 0.4, 0.6, 0.8, 1\}$.

As shown in \autoref{fig:ablation-model}, the performance of \sys is highly sensitive to both model selection and temperature ($T$). Lower temperatures (\eg, $T \leq 0.2$) maximize precision; notably, GPT-5 mini yields the fewest false positives by strictly adhering to known security patterns. However, these deterministic settings occasionally miss complex, multi-step vulnerabilities. 
In contrast, mid-range temperatures (\eg, $T \in [0.4, 0.6]$) improve recall by enabling exploratory reasoning, allowing Claude 3.7 Sonnet to uncover additional logical flaws at $T=0.4$ that were missed at $T=0$. Beyond $T=0.8$, performance degrades across all models due to increased hallucinations and the flagging of benign code.

\PP{Prompting Strategies}
By default, the \sys prompt includes two in-context examples (few-shot) and a structured three-step workflow to facilitate Chain-of-Thought (CoT) reasoning.
To isolate the impact of these components, we evaluate two additional strategies. First, we test \textit{zero-shot prompting}, where we remove the two demonstration examples to assess the model's reliance on in-context learning and its ability to identify vulnerabilities from its pre-training data.
Second, we evaluate \textit{direct prompting (No-CoT)}, where we remove the three-step reasoning instructions and require the model to output the final results directly. 
The same in-context examples are used.

\begin{table}[h]
\centering
\caption{Ablation of prompting strategies on Claude 3.7 Sonnet ($T=0.2$), evaluated on the ground-truth dataset.}
\label{tab:prompt-impact}
\begin{tabular}{lcccc}
\toprule
Strategy & TP & FP & Prec. & Recall \\ \midrule
\sys (Few-shot + CoT) & 17 & 4 & 81.0\% & 85.0\% \\
Zero-shot             & 14 & 6 & 70.0\% & 70.0\% \\
Direct (No-CoT)       & 11 & 9 & 55.0\% & 55.0\% \\ 
\bottomrule
\end{tabular}
\end{table}

As shown in \autoref{tab:prompt-impact}, the default \sys configuration achieves the highest performance, recovering 17 of the 20 vulnerabilities (85.0\% recall) on this ground-truth dataset. Removing few-shot demonstrations reduces the model's ability to ground its analysis, leading to a drop in both precision and recall. Furthermore, removing the CoT reasoning workflow results in the most significant degradation, confirming that intermediate Chain-of-Thought steps are essential for tracing complex data flows and minimizing false positives.

\summarybox{
Application-specific privileged operations increase recall from 33.3\% to 92.9\%, while code search primitives enable flexible search and avoid failures.
The validation strategies together eliminate 62.2\% of false positives.
}

\subsection{Comparison}

In this section, we compare \sys with prior classic and agentic state-of-the-art program analysis approaches on our dataset.
Ideally, a comparison would require labeling all vulnerabilities in the evaluated applications with ground truth.
However, this would require significant manual effort and is challenging (if not infeasible) to ensure completeness.
Therefore, we follow a common practice~\cite{mscan, witcher} and construct a consolidated comparison set as the union of vulnerabilities identified by all evaluated baselines and \sys.
    In total, this \emph{consolidated comparison set} consists of 44 vulnerabilities, including 42 vulnerabilities described in \autoref{s:eval-effectiveness} and 2 additional vulnerabilities identified by a baseline~\cite{enigma}.

\subsubsection{Baselines}

We compared \sys against three state-of-the-art approaches: CodeQL~\cite{codeql}, MScan~\cite{mscan}, and \enigma~\cite{enigma}.

\PP{MScan}
MScan~\cite{mscan} is a state-of-the-art static data-flow analysis tool designed to detect taint-style vulnerabilities in Java-based microservices.
While MScan was not originally designed for privilege escalation detection, it represents the closest comparable approach in the microservice security domain, as it performs taint analysis to identify flows to security-critical operations and supports cross-service interaction analysis.
To enable a meaningful comparison, we integrate \sys's validation components to adapt MScan for privilege escalation detection, denoted as MScan*.
The key difference between vanilla MScan and MScan* is that the latter incorporates \sys's permission check validation, while vanilla MScan only performs taint flow analysis without validating whether appropriate authorization checks exist.
We further build MScan*+Sinks, which extends MScan* with \sys's identified application-specific privileged operations, replacing MScan's default set of common security-critical operations.
Due to MScan's language limitation, we evaluate it only on the Java-based microservices in our dataset.
To ensure a rigorous and fair comparison, we applied \sys’s post-processing validation to the raw outputs of MScan* and MScan*+Sinks. This filters common static analysis noise, allowing for a direct comparison of each tool's capability against \sys.

\PP{CodeQL}
CodeQL~\cite{codeql} is a widely used static analysis tool that we include as a baseline for vulnerability detection.
It supports all programming languages in our dataset through its multi-language analysis framework.
However, CodeQL lacks inter-service analysis capabilities and treats each service independently, making it unable to detect cross-service privilege escalation vulnerabilities spanning multiple microservices.
We run CodeQL's default security query suite in the critical category and apply it to each service individually.
Similarly, we integrate \sys's validation components with CodeQL's results, denoted as CodeQL*.

\PP{\enigma}
\enigma~\cite{enigma} is an LLM-powered security testing agent developed based on SWE-agent~\cite{sweagent}, which leverages LLMs to automatically discover vulnerabilities through dynamic code exploration and reasoning.
As introduced in \autoref{s:background-agent}, \enigma relies on basic bash commands (\eg, \cc{grep}, \cc{find}) to navigate and read code files.
It loads code into the LLM's context window to reason about data flows and security implications directly through natural language understanding.
For meticulous whole-project analysis, we use the same setup as \sys.
Specifically, we provide the same standardized prompt strategy used by \sys (\autoref{s:impl}), together with the application’s documentation.
The same time budget of 10 hours is enforced for running \enigma.

\begin{figure}[t]
    \centering
    \begin{subfigure}[b]{0.4\columnwidth}
        \centering
        \includegraphics[width=\linewidth]{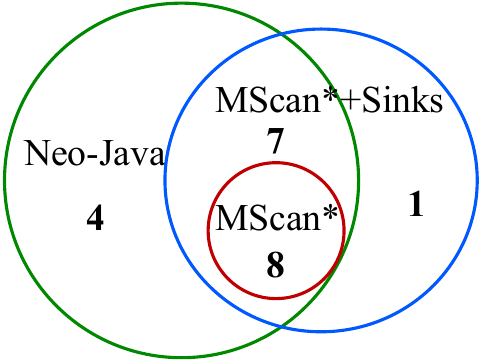}
        \caption{MScan.}
        \label{fig:venn-mscan}
    \end{subfigure}
    \hfill
    \begin{subfigure}[b]{0.4\columnwidth}
        \centering
        \includegraphics[width=\linewidth]{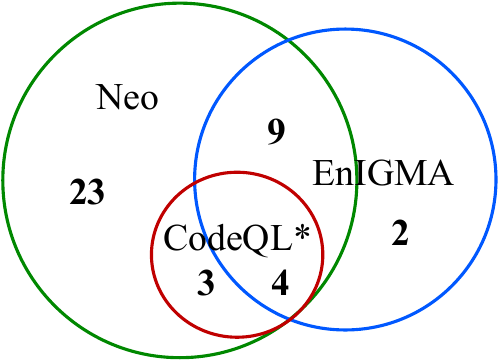}
        \caption{CodeQL and \enigma.}
        \label{fig:venn-other}
    \end{subfigure}
\caption{Distribution of true positive detections by each approach. MScan and CodeQL are integrated with \sys's validation and marked with a suffix *. MScan* is evaluated on Java-only vulnerabilities, while all other approaches are evaluated on the full polyglot dataset. MScan*+Sinks denotes MScan* enhanced with \sys's identified privileged operations.}
    \label{fig:venn}
\end{figure}

\begin{table}[t]
\centering
\scriptsize
\caption{TP, FP, and precision of compared tools.
}
\label{tab:precision}
{
\begin{tabular}{cccccc} 
\toprule
& MScan* & MScan*+Sinks & CodeQL* & \enigma & \sys \\
\midrule
TP & 8 & 16 & 7 & 15 & 39 \\
FP & 3 & 7 & 6 & 14 & 12 \\
Prec. & 72.7\% & 69.6\% & 53.8\% & 51.7\% & 76.5\% \\
\bottomrule
\end{tabular} 
}
\end{table}

\subsubsection{MScan Comparison}
Overall, MScan* found 8 privilege escalation vulnerabilities while MScan*+Sinks found 16, with the additional 8 vulnerabilities benefiting from the application-specific privileged operations identified by \sys.
In comparison, \sys found 19 vulnerabilities on the same Java-based microservices.
We present the distribution of true positives found by each tool in the Venn diagram in \autoref{fig:venn-mscan}.
The majority of vulnerabilities detected by MScan* or MScan*+Sinks were also found by \sys.
The only exception is CVE-2023-38493~\cite{CVE-2023-38493}, which was detected by MScan*+Sinks but not by \sys, as explained earlier in \autoref{s:eval-effectiveness}.
This occurred because MScan*+Sinks' (or MScan*'s) path analysis considered Spring's matrix variable normalization behavior, whereas \sys did not model this framework-specific URL processing semantic.
MScan* and MScan*+Sinks reported 3 and 7 false positives, leading to a precision of 72.7\% and 69.6\%, respectively.

\subsubsection{CodeQL Comparison}
CodeQL* detected 7 privilege escalation vulnerabilities, significantly fewer than \sys's 39 total detections across all languages.
As shown in \autoref{fig:venn-other}, all vulnerabilities found by CodeQL* were also covered by \sys.
This demonstrates the superior detection capability of \sys, which benefits from cross-service analysis and application-specific privileged operation detection.
CodeQL* achieved a precision of 53.8\% with 6 false positives.

\subsubsection{\enigma Comparison}

\enigma detected 15 privilege escalation vulnerabilities out of the 44 in the consolidated comparison set with 14 false positives.
As shown in \autoref{fig:venn-other}, \sys detected 26 vulnerabilities that \enigma missed, demonstrating superior detection coverage.
\enigma’s failures are primarily due to context exhaustion and the inefficiency of bash-based exploration. 
For example, in DeathStarBench~\cite{DeathStarBench}, a single data-flow path could span over 35 functions across different services.
Resolving the flows via bash commands returned massive irrelevant code that overwhelms \enigma to eventually give up the analysis. \sys succeeded because search primitives could directly retrieve context relevant to the vulnerability path.
This demonstrates \sys's advantage in static guidance, which filters out noise across large codebases.

However, \enigma found 2 vulnerabilities that \sys missed.
We analyzed \enigma's execution trajectories when interacting with the LLM to understand how it discovered these vulnerabilities and why \sys missed them.
In both cases, \sys successfully identified the privileged operations but failed to construct complete data flows due to complex dynamic function calls in JavaScript that its flow query operations could not track.
Consequently, these flows never reached \sys's LLM component for further analysis.
In contrast, \enigma directly reads and reasons about code through natural language understanding, without relying on program analysis to construct data flows.
This allows \enigma a chance to identify flows obscured by dynamic language features.

\summarybox{
\sys detected 39 of 44 vulnerabilities, substantially outperforming MScan, \enigma, and CodeQL.
This can be attributed to its LLM-based privileged operation identification and code search primitives.
}

\subsection{Efficiency and Cost}
\label{s:eval-cost}
We measured the analysis time and API cost of \sys when conducting the experiments.
\sys took a total of 38.9 hours to complete the analysis of all 25 applications, averaging 1.6 hours per application.
This end-to-end time includes privileged operation identification, cross-service flow analysis, on-demand context retrieval, LLM invocations, and final validation.
The average API cost was approximately 18 USD per application.
Given the complexity and scale of the microservices in our dataset, the efficiency and cost are practical for real-world deployment.
As a comparison, on average, \enigma took 3.4 hours and around 25 USD per application.
Note that the API cost of \enigma varied significantly across applications.
It frequently triggered early termination or, conversely, led to repeated context exhaustion.

\subsection{Extensibility}
\label{s:eval-usability}

\begin{table}[t]
\centering
\scriptsize
\caption{New vulnerabilities found in our extensibility study.}
\label{tab:more-vuls}
{
\begin{tabular}{lccc} 
\toprule
Vulnerability Type & TP & Ack'ed & Fixed\\
\midrule
Privilege Escalation & 11 & 5 & 2\\
Command Injection & 5 & 2 & 0 \\
SQL Injection & 2 & 0 & 0 \\
\bottomrule
\end{tabular} 
}
\end{table}

The underlying design of \sys is not limited to privilege escalation detection in microservices.
We argue that \sys has broad applicability across different contexts with only minor modifications.
To validate this claim, we apply \sys to detect privilege escalation in generic applications and to detect other types of vulnerabilities.

\PP{Adaptability to Generic Applications}
By design, the code search primitives and agentic program analysis pipeline in \sys are not specific to microservices.
The core techniques, such as code search and flow analysis, remain applicable to generic applications.
To validate how \sys adapts to generic applications, we modified the prompts to specify appropriate program context rather than assuming microservice architecture, and applied the modified \sys to several real-world open-source Python and Java applications.
As shown in \autoref{tab:more-vuls}, the modified \sys successfully identified 11 new privilege escalation vulnerabilities in these generic applications, with 5 acknowledged and 2 fixed by developers.
These results demonstrate that \sys's techniques generalize beyond microservices, confirming that the code search primitives and agentic analysis pipeline are broadly applicable.

\PP{Extensibility to Other Vulnerability Types}
While \sys includes specialized components for privilege escalation (\eg, privileged operation identification and authN/authZ detection), its core architecture generalizes to any taint-style, data-flow-based vulnerability.
The source-to-sink analysis paradigm naturally applies to SQL injection, XSS, command injection, and similar vulnerability classes that are prevalent in web and cloud applications.
Adapting \sys requires only redefining sources, sinks, and sanitizers through prompts, and the analysis pipeline remains unchanged.

We configured \sys to detect injection-based vulnerabilities~\cite{witcher} via prompt modifications.
As shown in \autoref{tab:more-vuls}, \sys identified 5 command injection and 2 SQL injection vulnerabilities in JavaScript-based web applications and packages, with 2 acknowledged by developers.
This confirms \sys's extensibility to diverse vulnerability types without architectural changes.

\subsection{Case Study}
\label{s:eval-case-study}
We present a case study to illustrate how \sys identified a privilege escalation vulnerability in Newbee Mall~\cite{newbeemall, newbeemall1}, a popular e-commerce application with 11.4K stars on GitHub.
As shown in \autoref{code:newbee}, the vulnerability exists in a payment endpoint where the \cc{paySuccess} function allows users to mark orders as paid by providing an order number.
While it validates that the order is in prepaid status (\cc{ORDER\_PRE\_PAY}), it lacks authorization to verify order ownership.
Any authenticated user can thus mark arbitrary orders as paid without actual payment.

\sys identifies this vulnerability through its multi-stage pipeline.
First, \sys identifies payment update operations as privileged by searching code with the function name \cc{order} using $Q_{ast}$ and $Q_{name}$, and the returned function calls are then validated by the LLM, which identifies lines 11-13 as privileged operations.
Second, code search primitives trace the flow from user-controlled endpoints to the identified privileged operations.
Third, the LLM-based validator explores the code to detect possible authN/authZ checks and identifies the missing ownership check, then further assesses its security impact.
During this process, on-demand context retrieval examines other similar methods in the file, such as \cc{cancelOrder()}, and discovers that they properly implement ownership verification (\eg, \cc{order.getUserId().equals(currentUser.getId())}), confirming this is a genuine vulnerability.

\begin{figure}[t]
    \inputminted[breaklines=true, frame=single]{Java}{code/newbee.java.tex}
    \caption{A (simplified) privilege escalation in Newbee Mall that allows setting an arbitrary order as paid.}
    \label{code:newbee}
\end{figure}

\subsection{Responsible Disclosure}
\label{s:eval-disclosure}

We follow standard responsible disclosure practices to protect users and developers from potential harm.
For each discovered vulnerability, we notified the corresponding developers or project maintainers.
As of this submission, 18 out of \numnewvultotal new vulnerabilities from \autoref{s:eval-effectiveness} and \autoref{s:eval-usability} have been acknowledged by developers, and 8 have been patched.
We continue to monitor the status of the remaining reports and communicate with developers to ensure timely fixes.

\section{Discussion}
\label{s:discussion}

\PP{Limitations}
We acknowledge a few limitations of \sys's current implementation.
First, \sys's code search primitives are built on CodeQL and inherit its limitations in handling dynamic language features (\eg, dynamic function calls and reflection) and pointer aliases.
While \sys supports multiple programming languages, its effectiveness is also restricted by CodeQL's language support and the quality of language-specific models.
One can also prototype the idea of code search primitives using other analysis backends such as Joern~\cite{joern}.
Second, \sys's code analysis component focuses on source code and cannot directly process configuration files (\eg, YAML, JSON, XML) that may contain relevant security settings such as access control policies or service routing rules.
Therefore, \sys relies on LLMs to interpret configuration content when retrieved as context, rather than performing structured semantic analysis.
Vulnerabilities stemming from such misconfigurations may be missed if the relevant configuration files are not retrieved into context.
Furthermore, many misconfigurations often depend on real-world deployment environments, which are typically inaccessible in a third-party red-teaming scenario. 
Given these constraints, privilege escalation via misconfiguration is out of scope for \sys.
Interested readers may refer to dedicated tools like ConfigX~\cite{zhang2021static} for detecting them.

\PP{Future Work}
Beyond addressing the aforementioned limitations, several promising directions could extend \sys's capabilities and impact. 
First, to mitigate the false positives, such as the ones identified in \autoref{s:eval-effectiveness}, we plan to refine the LLM's context-awareness to better distinguish execution environments, such as identifying logic unique to client-side components. 
Furthermore, we intend to explore hybrid analysis approaches that combine static analysis with dynamic testing and under-constrained symbolic execution.
This could enable \sys to navigate complex authN/authZ mechanisms and facilitate the automatic generation of proof-of-concept exploits to verify vulnerabilities. Finally, we plan to investigate how incremental analysis techniques could enable \sys to efficiently analyze code changes in continuous integration pipelines, making it more practical for ongoing development workflows.

\section{Related Work}
\label{relw}

\PP{Program Analysis for Microservices}
MScan~\cite{mscan} performs cross-service data-flow analysis for taint-style vulnerabilities in Java-based microservice applications.
It does not model or consider permissions or access control for privilege escalations.
ANTaint~\cite{wang2020scaling} builds an efficient and sound call graph and handles taint propagation in libraries by specifying shortcut rules over function arguments and return values.
Although both techniques target Java applications, their ideas are complementary to \sys.
For example, \sys currently builds on CodeQL for multi-language support, which also constrains it by CodeQL’s precision in call-target resolution and the resulting call graph quality.
Microscope~\cite{microscope} performs change-impact analysis to identify how code changes in one service propagate to others.
It models polyglot microservices using a unified Datalog representation and defines Datalog rules to determine the set of impacted public interfaces.
In contrast, \sys conducts a more fine-grained cross-service data-flow analysis that explicitly tracks the propagation path from user inputs to privileged operations, enabling the detection of security vulnerabilities.
AUTOARMOR~\cite{mspolicy} statically analyzes microservice source code to extract inter-service access control policies that define which services are authorized to make network requests to others.

\PP{Privilege Escalation Detection}
An open challenge in this domain is inferring the intended security policy for privileged operations.
Prior code analysis work often employs heuristics to approximate these intended policies~\cite{mace, son2011rolecast, pcfinder, mocguard, mpchecker}.
MACE~\cite{mace} computes authorization contexts (comprising users, roles, session variables, and permissions associated with sensitive operations) and compares these contexts across related operations (\eg, database \cc{INSERT} and \cc{DELETE}).
It flags inconsistencies as potential cases of missing authorization checks.
RoleCast~\cite{son2011rolecast} further utilizes knowledge of user roles and their associated permissions to detect inconsistencies.
Another line of work focuses on identifying existing permission checks and flagging cases where such checks are missing.
PCFinder~\cite{pcfinder}, for instance, leverages naming conventions of credentials (\eg, variables such as \cc{password}) and other semantic cues to detect permission checks.
MOCGuard~\cite{mocguard} analyzes database operations to infer data ownership and identifies missing owner-checks.
MPChecker~\cite{mpchecker} refers to the application log mechanisms to infer privileged operations that necessitate permission checks.
\sys approaches the open challenge in the context of polyglot microservices by leveraging LLMs for policy inference.

\PP{LLM-based Program Analysis}
Leveraging LLMs for program analysis has become an emerging trend.
RepoAudit~\cite{repoaudit} employs LLMs to navigate large codebases and perform on-demand analyses.
Recent research also explores agentic and hybrid techniques that combine LLM reasoning with traditional static or dynamic analyses.
IRIS~\cite{iris}, for instance, uses LLMs to infer taint sources and sinks and augments existing CodeQL queries with the inferred information.
LLMxCPG~\cite{llmxcpg} constructs program slices and leverages an LLM to validate them, while LLMSA~\cite{llmsa} decomposes analysis tasks into smaller property checks for improved accuracy and interpretability.
LLift~\cite{llift} applies LLM reasoning to infer post-conditions for use-before-initialization bugs in the Linux kernel.
In contrast, \sys emphasizes the importance and necessity of rigorous program analysis to support scalable and efficient analysis, which could not be handled with only LLMs.

\section{Conclusion}
\label{s:conclusion}
In this work, we presented \sys, an LLM-based agentic program analysis framework for detecting privilege escalation vulnerabilities in polyglot microservices.
\sys leverages LLMs to reason about natural language semantics and employs rigorous program analysis to analyze structured code.
Our carefully designed code search primitives enable \sys to flexibly and scalably explore large codebases. 
On real-world microservice applications, our evaluation uncovered \numnewvul zero-day privilege escalation vulnerabilities.
\sys significantly outperformed existing approaches in detection accuracy, scalability, and extensibility.
We believe such an agentic program analysis pipeline represents a promising paradigm for addressing emerging software security challenges in the long term.

\section*{Ethics Considerations}

We identify application owners and their end users as the primary stakeholders that may potentially be affected by our work. 
To minimize potential risks and harms, we adopted a proactive, responsible disclosure approach by reporting newly identified vulnerabilities immediately upon discovery.
Our evaluation spanned two months, from September to October 2025, during which we reported findings to affected software vendors and strongly encouraged prompt remediation.
To prevent in-the-wild exploitation, we did not publicly disclose vulnerability details for any reported issues prior to vendor patches being made available.
As of this submission, 18 out of \numnewvultotal vulnerabilities have been acknowledged by developers, and 8 have been patched.

All experiments, vulnerability validation, and proof-of-concept testing were conducted exclusively in isolated local environments under our control.
We reconstructed vulnerable microservice configurations locally to reproduce and verify privilege escalation paths without interacting with production systems or accessing real user data.
Our methodology ensured no live deployments were compromised and no real-world service disruptions were caused.

\sys employed LLMs to analyze application code and identify privilege escalation vulnerabilities.
To ensure responsible use, all LLM analysis was conducted in isolated environments and excluded sensitive credentials, proprietary code, and personally identifiable information.
Final LLM outputs were manually validated by security researchers before reporting to vendors.

We acknowledge that \sys, as a vulnerability detection framework, could potentially be extended or misused for malicious purposes.
To mitigate this risk, we release our artifact in a research-oriented form, excluding specific vulnerability details and exploit payloads.
We believe the societal benefits of this work significantly outweigh the limited risks.

\section*{LLM Usage Considerations}

LLMs were used for editorial purposes in this manuscript, and all outputs were inspected by the authors to ensure accuracy and originality. The literature review, \sys framework design, system implementation, and experimental evaluation were all conducted independently by the authors without LLM assistance. During the writing process, we employed LLMs to polish the manuscript by improving sentence structure, correcting grammar, and enhancing the clarity of technical explanations. All substantive content, including the methodological approach, experimental findings, security analysis, and research conclusions, represents the authors' original contributions. We carefully reviewed all LLM-generated editorial suggestions to verify their accuracy and appropriateness before incorporation.

LLMs are used in \sys's design for vulnerability detection. \sys leverages Claude Sonnet 3.7 as an agentic reasoning component that dynamically generates analysis plans, adapts code search strategies, and validates security semantics. All LLM-generated analysis decisions were executed in controlled environments for analyzing open-source microservice applications. Our use of a closed-source model may introduce reproducibility challenges. To mitigate that, we document exact model versions and parameters, and provide major prompts used in the framework in the artifact. We manually verified all detection results to confirm their validity and responsibly disclosed the vulnerabilities to the relevant developers for remediation. We present all detection statistics in the paper. 

We do not train any models in this work. \sys analyzes open-source microservice applications from publicly available GitHub repositories, which do not raise concerns regarding consent, data holder rights, or intellectual property. Our environmental footprint is limited to inference costs totaling 38.9 hours of LLM API usage, as discussed in Section 6.5. This cost is justified by \sys's goal of automating privilege escalation detection in complex polyglot microservices, a task requiring extensive manual security expertise. We minimized our footprint by using code search primitives to retrieve only relevant code snippets instead of reading entire codebases.

\section*{Acknowledgments}
\label{s:ack}
The authors would like to thank the anonymous reviewers and shepherd for their constructive comments.
We also thank Andrew Johnston for the valuable discussions.
This work was supported in part by the National Science Foundation (NSF) under grants CNS-21-54404 and CNS-20-46361, Gifts from Google, the Columbia Research Stabilization Fund, and the Columbia OPA Academic Conference Travel Grant.
The views and conclusions contained herein are those of the authors and should not be interpreted as necessarily representing the official policies or endorsements, either expressed or implied, of the funding agencies.

\bibliographystyle{IEEEtran}

\bibliography{p, conf}
\appendices
\onecolumn \begin{multicols}{2}

\twocolumn
\section{Meta-Review}

The following meta-review was prepared by the program committee for the 2026
IEEE Symposium on Security and Privacy (S\&P) as part of the review process as
detailed in the call for papers.

\subsection{Summary}
This paper presents \sys, an agentic program analysis framework for detecting privilege escalation vulnerabilities in microservice architectures. \sys combines the semantic reasoning capabilities of LLMs with the traditional static program analysis. \sys uses an LLM-driven agent to iteratively plan and perform code queries across services and languages, identifying privileged operations, tracing cross-service data flows, and checking for missing or inadequate authentication. Evaluated on 25 open-source microservice applications, \sys finds 24 vulnerabilities. \sys significantly improves detection coverage and reduces missed vulnerabilities, compared to existing tools.

\subsection{Scientific Contributions}
\begin{itemize}
\item 3. Creates a New Tool to Enable Future Science.
\item 4. Addresses a Long-Known Issue.
\item 5. Identifies an Impactful Vulnerability.
\item 6. Provides a Valuable Step Forward in an Established Field.
\end{itemize}

\subsection{Reasons for Acceptance}
\begin{enumerate}
\item The paper presents a well‑designed system that effectively integrates LLMs with static analysis, enabling scalable and targeted detection of privilege‑escalation vulnerabilities in complex microservice architectures. \sys's use of structural code‑search primitives avoids the limitations of large‑context ingestion and allows the agent to reason over heterogeneous, cross‑service codebases.
\item The evaluation is strong and comprehensive. This paper analyzed 25 real‑world microservice applications, uncovering previously unknown vulnerabilities---many confirmed and patched by developers---demonstrating clear practical impact. The system also generalizes beyond its primary task, detecting other bug classes such as command injection and SQL injection with only minor prompting changes, indicating broader applicability.
\item Architecturally, \sys is well‑motivated and logically structured, and its design is validated by robust experiments, diverse baselines, and well‑posed research questions. The inclusion of runtime and API cost measurements adds transparency into the system’s operational overhead.
\end{enumerate}

\subsection{Noteworthy Concerns} %
\begin{enumerate} %
\item \textit{Model and Prompt Sensitivity and Reproducibility Concerns}: \sys relies heavily on LLM reasoning and prompt design, as a result, using different LLM models may impact the results and require additional prompt tuning.

\item 
\textit{Dependence on CodeQL and Its Limitations}: \sys heavily depends on CodeQL, inheriting its constraints and reducing portability to other analysis backends.

\end{enumerate}

\end{multicols}
\end{document}